\documentclass[twocolumn,prl]{revtex4}
\usepackage{graphicx}
\usepackage{subfigure}
\begin{document}

\title{Self-Polarization and Dynamical Cooling of Nuclear Spins in Double Quantum Dots}
\author{M. S. Rudner, L. S. Levitov}
\affiliation{
 Department of Physics,
 Massachusetts Institute of Technology, 77 Massachusetts Ave,
 Cambridge, MA 02139}

\date{\today}

\begin{abstract}
The spin-blockade regime of double quantum dots features coupled dynamics of electron and nuclear spins resulting from the hyperfine interaction.
We explain observed nuclear self-polarization via a mechanism based on feedback of the Overhauser shift on electron energy levels, and propose to use the instability toward self-polarization as a vehicle for controlling the nuclear spin distribution.
In the dynamics induced by a properly chosen time-dependent magnetic field, nuclear spin fluctuations can be suppressed significantly below the thermal level. 
\end{abstract}

\maketitle

Recent advances in semiconductor quantum dot technology have given experimentalists the ability to control the behavior of individual electrons and investigate their coupling to nuclear spins.
Interesting phenomena such as switching, hysteresis, and long period oscillatory behavior of electric current were observed by Ono and Tarucha\cite{OnoTarucha} in the so-called spin-blockade regime in GaAs vertical double quantum dots.
Koppens et al. \cite{Koppens} have also observed bistability and switching in a lateral quantum dot system.
In both cases, strong evidence was presented linking the observed phenomena to collective behavior of the nuclear spins in the lattice.

Because of the coupling of electron and nuclear spins, uncertainty in the nuclear spin state leads to undesirable effects in electron spin dynamics such as dephasing\cite{Petta} and fluctuations in Zeeman energy\cite{ESR}.
Learning how to control nuclear spins will open up new possibilities for nuclear spin-based information storage and manipulation, and improve our ability to coherently control the behavior of electron spins.

Ordering of nuclear spins is possible in equilibrium only at microkelvin temperatures, due to the weakness of the dipole-dipole interaction between nuclear spins.
The instabilities observed at Kelvin temperatures in \cite{OnoTarucha, Koppens}, however, indicate spontaneous ordering under more accessible conditions.
Here we explain these observations and propose to use the dynamics of this hysteretic regime to narrow the distribution of nuclear spin polarization.

Much of the recent theoretical work on spins in quantum dots has focused on dephasing \cite{KLG, CL} and relaxation \cite{ENF} of electron spins due to their interaction with the lattice nuclei.
The interplay of nuclear spin dynamics and spin-blockaded electron transport has also been studied \cite{JN, MacDonald}.

In this Letter we are interested in self-polarization as a means to control the behavior 
of nuclear spins.
We present a model of electron transport through a double quantum dot system which exhibits an instability toward self-polarization of the nuclei.
The ideas of feedback and self-polarization date back more than 30 years to the work of Overhauser~\cite{Overhauser} and of Dyakonov and Perel~\cite{DP}.
Here we extend these ideas to the interesting physics relevant to spin-blockaded quantum dots.
We identify an additional essential ingredient that is needed to achieve polarization: electrons with one spin orientation must prefer to exchange spin with a nucleus to escape, while electrons with opposite spin escape primarily by another means without exchanging spin with the nuclear system.

We discuss the conditions that control self-polarization and analyze the width of the nuclear spin distribution.
The restoring force near the polarized steady-state turns out to be weak and does not help to suppress fluctuations.
However, we find that the dynamics in other regimes of polarization possess a squeezing property that tends to narrow the nuclear spin distribution.
We propose a scheme to harness this squeezing effect by applying a time-dependent external magnetic field.
Estimates indicate that fluctuations can then be suppressed significantly below the thermal level.

We start with reviewing the model put forth in Ref.\cite{OnoTarucha} to explain the behavior of vertical double quantum dot devices in the spin blockade regime.
This model, accepted here for concreteness and to set the stage for our discussion, illustrates more general ideas that apply to a variety of systems with similar electron energy spectra.

\begin{figure}
\includegraphics[width=3.0in]{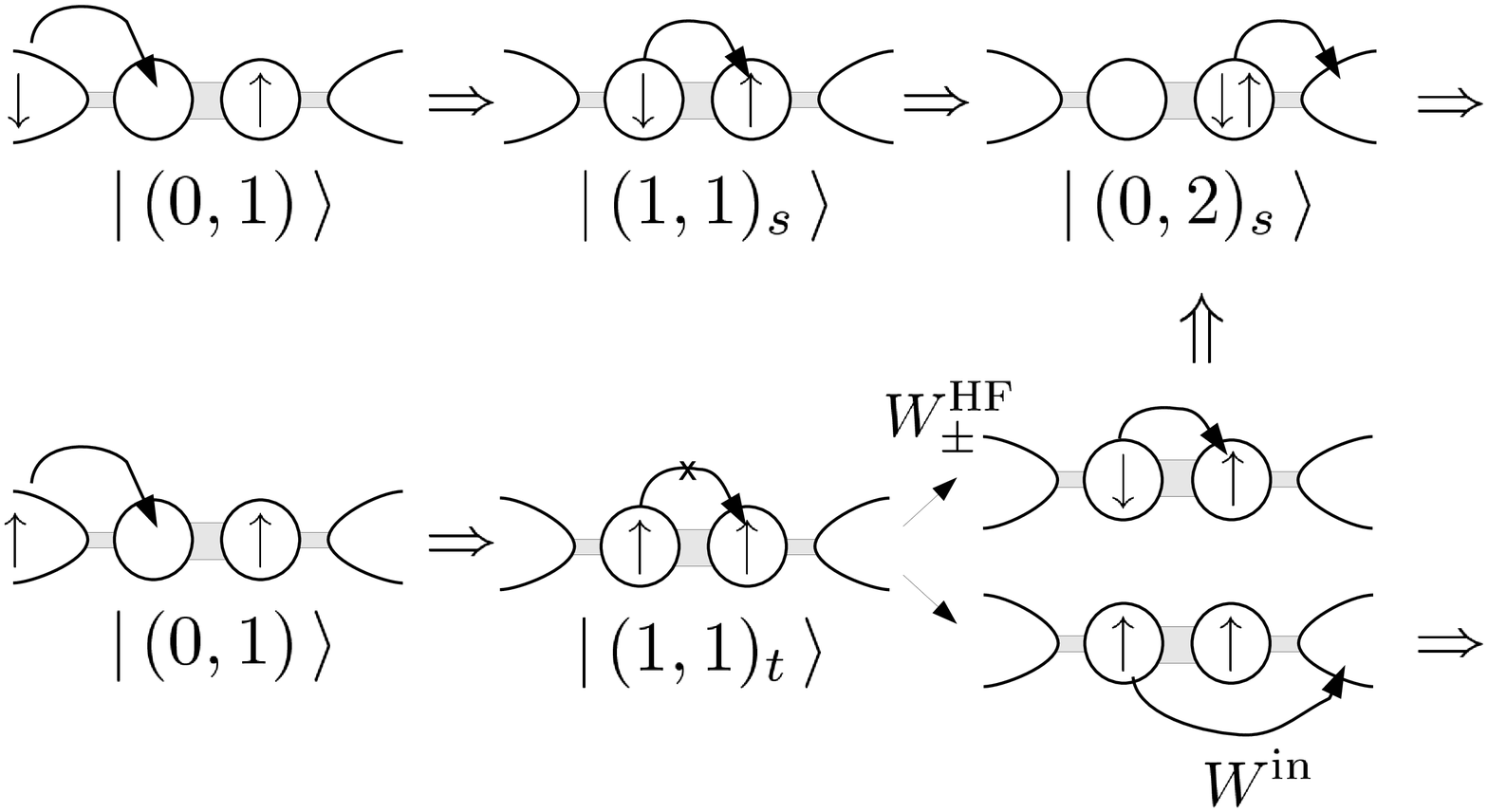}
\vspace{0.15cm}
\vspace{-3.5mm}
 \caption[]{Spin blockaded transport through the double quantum dot system.  Initially the system is in the state $(0,1)$ with one electron on the right dot.  
Current flows as depicted by the arrows: an electron tunnels from the source into the left dot to form the state $(1,1)_{s/t}$ with singlet or triplet spin, respectively.  
The electron then must hop to the right dot to form $(0,2)_s$ before tunneling out and returning the system to the $(0,1)$ state.  
If $(1,1)_t$ is formed (bottom), the Pauli Principle prohibits the second electron from tunneling into the right dot.  
The triplet can decay via hyperfine spin flip with rate $W_{\pm}^{\rm HF}$,
Eq.(\ref{eqnFGR}), or by indirect tunneling with rate $W^{\rm in}$.}
\label{fig:transitions}
\vspace{-.25in}
\end{figure}

The double dot is weakly coupled in series to two unpolarized leads, with transport occuring as a series of discrete hopping events as described in Figure \ref{fig:transitions}.
Current is suppressed when one of the three $(1,1)_t$ states is occupied in the first step of the cycle.  
Once in the triplet state, residual current can result from slow indirect tunneling through virtual excited states, exchange with the leads, and spin-flips due to spin-orbit coupling and/or hyperfine flip-flop scattering.  
Because spin-orbit effects are suppressed due to confinement in structures of this type~\cite{KN, GKL, Elzerman, Amasha}, we consider only indirect tunneling and hyperfine scattering.

For simplicity, we assume that tunneling rates from the weakly coupled source lead into the four $(1,1)_{s/t}$ states are all equal.
Coupling of the $(0,2)_s$ state to the drain lead is strong and gives this level a large decay width. 
The finite width of this level and of $(1,1)_s$ due to hybridization/orbital relaxation plays a key role in the feedback mechanism that drives the polarization instability by favoring transitions from one of the split-triplet levels to $(1, 1)_s$ as they are brought into resonance. 

In general, the orbital eigenstates with singlet spin configuration are a superposition of the states $(1,1)_s$ and $(0,2)_s$.
Here we assume, for simplicity, that one eigenstate retains predominantly the $(1,1)$ charge distribution, while the other retains predominantly the $(0,2)$ charge distribution.
We refer to these ``$(1,1)_s$-like'' and ``$(0,2)_s$-like'' states in Fig. \ref{fig:transitions} and hereafter.

We assume incoherent nuclear dynamics, and describe the system by the populations $N_{\pm}$ of the up and down nuclear spin states \footnote{Although our primary example of GaAs is comprised of spin-3/2 nuclei, for clarity of presentation we proceed with the calculation assuming spin-1/2 nuclei. 
This allows us to introduce the populations of nuclei in the up and down spin states, $N_+$ and $N_-$.}.
We also neglect spatial variations in the nuclear spin population and transitions that do not change the net spin, which are inessential for our analysis.
The energy-dependent hyperfine spin flip transition rates $W_{\pm}^{{\rm HF}}$ are calculated using Fermi's Golden Rule:
\begin{eqnarray}
 \label{eqnFGR}
  W_{\pm}^{{\rm HF}} = \frac{2 \pi}{\hbar} \vert \langle \,(1,1)_s \, \vert \hat{H}^{{\rm HF}} \vert \, (1,1)_{t_{\pm}} \,\rangle\vert^2 \, N_{\mp} \, f(\varepsilon_{\pm})
\end{eqnarray}
where $f(\varepsilon)$ is the density of states for the singlet final state, and $\varepsilon_{\pm}$ is the energy difference between the singlet final state and the triplet state with z-projection $\pm 1$.
We assume a Lorentzian lineshape
\[
  f(\varepsilon) \propto \frac{\gamma}{\varepsilon^2 + \gamma^{2}}.
\]
to allow explicit calculation.  

When electrons are injected from an unpolarized source and every electron must exchange its spin with a nucleus to escape, no spin can be pumped into the nuclear spin system irrespective of the ratio of the rates for flipping nuclei up or down, $W_+^{\rm HF}/W_-^{\rm HF}$.
If, on the other hand, electrons have an alternative way to escape, it need not be the case that the same number of nuclei must flip their spins in each direction.  

For simplicity, we assume a single energy-independent indirect tunneling rate $W^{\rm in}$ for all three triplet states that relieves spin-blockade without interaction with the nuclei.  
Because of the competition between these processes, the net nuclear spin flip rates are given by
\begin{eqnarray}
  \label{eqnPumpRates}
   \Gamma_{\pm} = \frac{W_{\pm}^{{\rm HF}}}{W_{\pm}^{{\rm HF}} + W^{{\rm in}}}\frac{I}{4},
\end{eqnarray}
\noindent where $I$ is the total current through the system\footnote{The current can be found self-consistently from the rates of the individual processes. The exact form is not essential to our analysis.}.
If the energy dependent rates $W_{\pm}^{{\rm HF}}$ are not equal, it is possible for electron transport to be dominated by spin-flip processes for electrons of one spin type, and by indirect tunneling processes for electrons of the other type.

The nuclear polarization is in a steady state when the opposing spin flip rates $\Gamma_+$ and $\Gamma_-$ are equal.
Assuming no dependence of the orbital matrix elements on the electron spin z-projection, we have:
\begin{eqnarray}
   \label{eqnStat}
  f(\varepsilon_+)\, N_- = f(\varepsilon_-)\, N_+.
\end{eqnarray}
When $f(\varepsilon_+) \neq f(\varepsilon_-)$, there can be a nonzero nuclear spin polarization $s \equiv N_+ - N_-$ in the steady state even when electron Zeeman energy is negligible compared to the lattice temperature.

\begin{figure}
\includegraphics[width=2.5in]{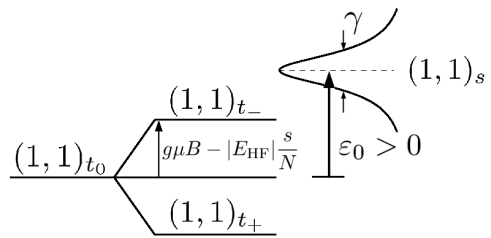}
\vspace{-3mm}
\caption[]{ Electron energy levels in the arrangement appropriate for GaAs, where $g\mu \approx 26 \, \mu {\rm eV}/{\rm T}$ and $E_{\rm HF} \approx -130 \, \mu {\rm eV}$.}
\label{figLevels}
\vspace{-5mm}
\end{figure}

Because of the hyperfine coupling induced Overhauser shift, the triplet state splitting acquires a polarization dependence in addition to the usual Zeeman splitting: 
\begin{eqnarray}
  \label{ESplitting}
  \varepsilon_{\pm} = \varepsilon_0 \pm \left[ g \mu B + E_{\rm HF} \cdot\left(s/N\right) \right],\ -N \le s \le N.
\end{eqnarray}
Here $\varepsilon_0$ is the singlet/triplet detuning as depicted in Figure \ref{figLevels}, $g \mu$ is the effective magnetic moment of the electron in the material, $B$ is the strength of the applied magnetic field, $E_{\rm HF}$ is the hyperfine energy for an electron localized on a single polarized nuclear spin (negative in GaAs), and $N$ is the total number of nuclear spins\footnote{Note that in GaAs $g$,  $\mu$, and $E_{\rm HF}$ are all negative.
Thus nuclear spins polarized in the {\it negative} z-direction add constructively to the applied Zeeman field.}.

Feedback in this system comes from the $s$ dependence of (\ref{ESplitting}), which leads to a polarization dependence of the spin flip rates $\Gamma_+$ and $\Gamma_-$.
Equations (\ref{eqnStat}) and (\ref{ESplitting}) together describe an instability of a similar form to that found by Dyakonov and Perel in their early work on self-polarization under optical pumping
\cite{DP}.

In the absence of an applied magnetic field, conditions (\ref{eqnStat}) and (\ref{ESplitting}) yield a third order equation for the equilibrium polarization $s^*$ with solutions
\begin{eqnarray}
   \label{eqnSolns}
   s^* = 0, \ \ s^{\ *}_{\pm} = \pm N\sqrt{2 \tilde{\varepsilon}_0 - (\tilde{\varepsilon}_0^{\ 2}  + \tilde{\gamma}^2)}.
\end{eqnarray}
where $\tilde{\varepsilon}_0 \equiv \varepsilon_0/\vert E_{\rm HF}\vert$ and $\tilde{\gamma} \equiv \gamma/\vert E_{\rm HF} \vert$.

Which of these solutions corresponds to a stable equilibrium?
The solution $s^* = 0$ always exists, for at $B = 0$ the triplet levels are degenerate when $s = 0$.
However, the solutions with finite nuclear polarization exist only when the discriminant is positive: $(1 - \tilde{\varepsilon}_0)^2 + \tilde{\gamma}^2 < 1$.
When this condition is met, the zero-polarization solution $s^* = 0$ is unstable.
The polarization (\ref{eqnSolns}) is maximized for the detuning from resonance $\varepsilon_0 = -E_{\rm HF}$, with $ s_{\rm max}=\pm N\sqrt{1-\tilde{\gamma}^2}$.

Because $E_{\rm HF}$ is negative in GaAs, the scaling by $\vert E_{\rm HF} \vert$ leads to the signs as shown in equation (\ref{eqnSolns}) and in the discriminant condition.
For materials with positive hyperfine energy, one should replace $\tilde{\varepsilon}_0$ with $-\tilde{\varepsilon}_0$ in these expressions.
In GaAs, the singlet level must be {\it above} the triplet for the zero field instability to occur, while in materials with opposite sign the singlet should be below.

Can the appearance of self-polarization result in a narrowing of the nuclear spin distribution?
The maximal polarization $s_{\rm max}$ found above scales as $N$ and can be arbitrarily close to $\pm N$.
If this were realistic, then one could indeed squeeze the nuclear spin distribution simply by allowing the system to become fully polarized.
In real systems, however, nuclear spin relaxation and diffusion will prevent complete polarization of the nuclear spins.

With this limit on achievable maximum polarization in mind, we now investigate the width of the nuclear spin distribution in the partially polarized steady state.
The rates given by equation (\ref{eqnPumpRates}) describe a model consisting of a series of electron tunneling events accompanied by stochastic unit steps in nuclear spin polarization.
Because the total number of nuclear spins is large, the resulting evolution of the nuclear spin distribution $\rho(s, t)$ is well described as a Fokker-Planck diffusion process characterized by a drift velocity $V = 2(\Gamma_+ - \Gamma_-)$ and diffusion parameter $D = 2(\Gamma_+ + \Gamma_-)$ \cite{VanKampen}:
\begin{eqnarray}
  \label{eqnFP}
  \frac{\partial}{\partial t}\rho(s, t) = \frac{\partial}{\partial s}\left(D(s) \, \frac{\partial}{\partial s}\rho(s, t) - V(s)\,\rho(s, t) \right).
\end{eqnarray}
In the steady state, the time-independent distribution $\rho_0(s)$ is given by $\rho_0(s) \propto \exp[\int_{s_0}^{s} V(s')/D(s') \, ds']$.

The peaks of this stationary distribution occur at the values of polarization corresponding to the stable fixed points described above.
Here we indeed see that a fixed point $V(s^*) = 0$ is stable if $dV/ds\vert_{s*} < 0$.
One can estimate the spread of the nuclear spin distribution about such a stable fixed point by linearizing the integrand above for $s$ near $s^*$.
The result is a Gaussian of width
\begin{eqnarray}
  \label{eqnFPWidth}
  \sigma_s = \sqrt{D/\left\vert dV/ds\right\vert}.
\end{eqnarray}

An important question to ask is how the width of this distribution compares with that of the randomized high temperature distribution.
For $N$ randomly oriented spins of unit length, one finds $\sigma_{\rm Th} = \sqrt{N/3}$.
Because $V$ depends on $s$ only through the combination $s/N$, expression (\ref{eqnFPWidth}) is also proportional to $\sqrt{N}$.
We find the prefactor to be relatively insensitive to parameters over a wide range, typically falling in the range $\sigma_s \approx (0.6-0.7)\sqrt{N}$.
The spontaneous polarization of nuclear spins by this mechanism thus does not narrow the distribution relative to the thermal state.

Why is this so?
Near the finite polarization fixed points, all hyperfine transitions are off-resonant and transport is dominated by indirect tunneling.  
Nuclear spin dynamics are thus slow, and do not provide a strong enough restoring force to squeeze the spin distribution below its thermal width.

It is possible, however, to achieve squeezing away from the steady state.
For that,
the derivative of the drift velocity $V$ with respect to polarization 
must be large and negative.  
In such a region, parts of the spin distribution that ``lag behind'' are pushed forward, while those that have ``run ahead'' are held back.

Rather than allowing the system to self-polarize, we now explore a way of dynamically trapping the system in a region of near maximal (negative) $dV/ds$.
Squeezing is most efficient when the energy of $(1, 1)_{t_-}$ is greater than that of $(1, 1)_s$ by approximately $\gamma$, i.e. midway down the upper shoulder of the resonance.
At zero polarization, this condition is met when $g\mu \, B = \varepsilon_0 + \gamma$.
However, in this resonant situation, a high rate of polarization drives the system away from the optimal squeezing regime.

Suppose that after a short time $\tau_{\rm rev}$ the direction of the external magnetic field is reversed. 
After reversal, the ensuing dynamics will tend to undo the polarization that develped during the first period.
By repeatedly reversing the magnetic field after consecutive periods of $\tau_{\rm rev}$, the polarization can be trapped in a range where the squeezing effect is always near maximal.

\begin{figure}
\includegraphics[width=3.25in]{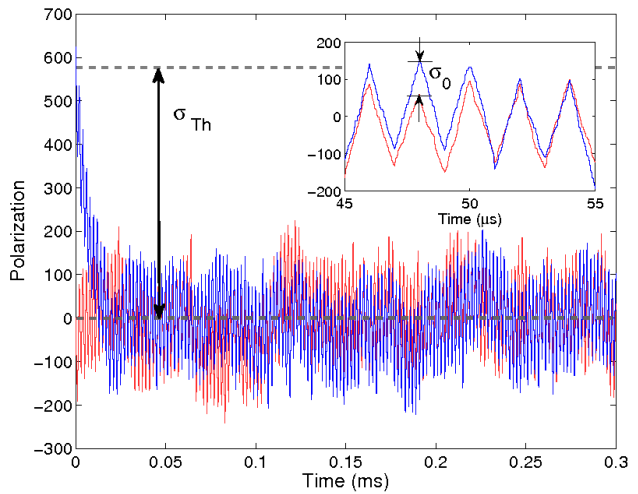}
\vspace{-0.15cm}
\caption[]{(color online). Squeezing of the nuclear spin distribution by periodic reversal of magnetic field.
Numerical results for two initial values of polarization (red and blue) with $N = 10^6$ spins and reversal time $\tau_{\rm rev} = 1 \,\mu {\rm s}$ are shown.  
Inset: A zoom-in on the interval $45 \, \mu {\rm s}$ to $55 \,\mu {\rm s}$.
The size of fluctuations about the limiting orbit, $\sigma_0$,
is well below the size of fluctuations
in the initial state, $\sigma_{\rm Th}$.
}
\label{BReversal}
\vspace{-5mm}
\end{figure}
We now examine squeezing in this model by considering the behavior of trajectories in the vicinity of a stable sawtoothlike limiting orbit of the dynamical system $\dot{s} = V(s, t)$ (the existence of such an orbit is easy to prove in the limit of fast switching).
Here the explicit time dependence of the drift velocity $V(s, t)$ arises from the time dependence of the external field.
For illustration (Fig. \ref{BReversal}), we choose parameter values to center the sawtooth around $s = 0$.

Through standard techniques of analyzing periodically driven systems, one finds that over one complete driving cycle ($\Delta t = 2 \tau_{\rm rev}$), the distance between nearby trajectories shrinks by the Lyapunov multiplier
\begin{eqnarray}
  \label{Lyapunov}
  \Lambda \approx 1 - \left\vert dV/ds\right\vert_{s=0} \Delta t,\ dV/ds < 0.
\end{eqnarray}

The stochastic nature of the dynamics can be reintroduced by extending the discrete map described by (\ref{Lyapunov}) to continuous time.
A new Fokker-Planck equation can then be written down to describe the approach of trajectories to the limiting orbit, with drift velocity $V_0(s) = (dV/ds)_{s = 0} s$ and diffusion constant $D_0 = D(s = 0)$.

The width of the resulting steady state distribution 
\begin{eqnarray}
  \label{eqnBrevWidth}
  \sigma_0 = \sqrt{D_0/\vert dV/ds\vert_{s=0}}
\end{eqnarray}
characterizes the size of fluctuations about the limiting orbit.
Squeezing is most effective in the regime of fast indirect tunneling $W^{\rm HF}_{\pm}/W^{\rm in} \ll 1$.
To lowest order in $W^{\rm HF}_{\pm}/W^{\rm in}$, we find $\sigma_0 \approx \sqrt{3 \tilde{\gamma} N}$.
In this regime for a resonance of width $0.1 \, \mu {\rm eV}$, we estimate that the nuclear spin distribution can be narrowed down to a width $\sigma_0 \approx 0.05 \sqrt{N}$.

Figure \ref{BReversal} illustrates our numerical simulations of the stochastic dynamics of the transport cycle described in Figure \ref{fig:transitions}.
An electron is first loaded into one of the four initial states with uniform probability.
Transitions are made out of this state after a random time distributed according to the microscopic rates $W_{\pm}^{\rm HF}$ and $W^{\rm in}$ described above plus a fast decay rate $W^{\rm s}$ for the singlet.
The nuclear polarization state is updated whenever a spin flip transition is made.

We found that the basin of attraction of the limiting orbit is larger than the thermal width of the distribution, marked by the dashed line in Fig. \ref{BReversal}.
Thus the attractor is strong enough to pull in orbits from the full range of probable initial conditions (e.g. the blue trajectory in Fig. \ref{BReversal}).

The parameters $N=10^6$ and $\tau_{\rm rev} = 1\,\mu{\rm s}$ were chosen for convenience.
To apply these results to different $N$, note that the rate $W^{\rm HF}_\pm$ scales as $1/N$; consequently the timescale for polarization scales as $N^2$.
Thus for a dot with $N = 10^7$ nuclear spins, a switching time $\tau_{\rm rev} = 0.1\, {\rm ms}$ would yield the same behavior.

For the purpose of illustration, we display results for a situation where the fluctuations are comparable in size to the amplitude of the sawtooth.
When $\tau_{\rm rev}$ is increased, the amplitude of the sawtooth grows larger but fluctuations about the sawtooth remain relatively unaffected.
For applications of cooling, however, it is the size of fluctuations about the instantaneous mean that is important (see inset of Fig. \ref{BReversal}).
One can stop the dynamics at any point on the sawtooth to leave behind a cooled nuclear spin system that will persist for times up to the diffusion/relaxation time.
As the estimates above show, a significant enhancement of electron coherence times may be possible through this squeezing scheme.

We have seen that in the spin-blockade regime in double quantum dots nuclear spins exhibit collective effects such as self-polarization, ordering, and squeezing.
This regime provides means of controlling nuclear spins, thus enabling the coherent manipulation of electron spins.

We benefitted from useful discussions with D. A. Abanin, M. I. Dyakonov, F. H. L. Koppens, J. R. Petta, S. Tarucha and L. M. K. Vandersypen, and support from the W.M. Keck Foundation.
M.R.'s research is supported by DOE CSGF, Grant No. DE-FG02-97ER25307.

\vspace{-5mm}


\end{document}